\documentclass[%
 reprint,
 amsmath,amssymb,
 aps,
]{revtex4-1}

\usepackage{graphicx}
\usepackage{dcolumn}
\usepackage{bm}
\usepackage[mathlines]{lineno}
\linenumbers\relax 

\nolinenumbers

\begin{document}

\preprint{}

\title{Heavy ion beam loss mechanisms at an electron-ion collider}

\author{Spencer R. Klein}
 \email{srklein@lbl.gov}
\affiliation{Lawrence Berkeley National Laboratory, Berkeley, CA, USA}

\date{\today}

\begin{abstract}

There are currently several proposals to build a high-luminosity electron-ion collider, to study the spin structure of matter and measure parton densities in heavy nuclei, and to search for gluon saturation and new phenomena like the colored glass condensate.  These measurements require operation with heavy-nuclei. We calculate the cross-sections for two important processes that will affect accelerator and detector operations: bound-free pair production, and Coulomb excitation of the nuclei.   Both of these reactions have large cross-sections, 28-56 mb, which can lead to beam ion losses, produce beams of particles with altered charge:mass ratio, and produce a large flux of neutrons in zero degree calorimeters.  The loss of beam particles limits the sustainable electron-ion luminosity to levels of several times $10^{32}/$cm$^2$/s.

\end{abstract}

\maketitle


Electron-ion colliders have been proposed as a means to study the structure functions of polarized protons and to probe the quark and gluon distributions of heavy nuclei \cite{Accardi:2012qut,AbelleiraFernandez:2012cc}.  The latter topic is of great interest in studying the behavior of quarks and gluons at high densities, such as are present in nuclei at low Bjorken-x values.  These studies require high electron-ion luminosities, so as to be able to probe reactions with low cross-sections, near the kinematic limit in $x$ and $Q^2$. 

At very high luminosities, other reactions, with large cross-sections, may occur copiously enough to cause significant beam loss.  Two such reactions are Coulomb excitation of heavy nuclei, and bound-free pair production (BFPP).  In BFPP, an electron-positron pair is produced, with the electron bound to the target nucleus.    BFPP leads to a single-electron ion, while Coulomb excitation leads to neutron emission and/or nuclear breakup.

Most Coulomb excitation occurs at low photon energies.  A nucleus is excited, typically to a Giant Dipole Resonance (GDR);  the GDR decays usually decays via single neutron emission, leaving a slightly lighter ion, plus a neutron
\cite{Baltz:1996as}.  At higher energies, Coulomb excitation can involve photon-nucleon interactions, such as excitation to a $\Delta$ resonance, or more energetic photon-quark interactions which lead to nuclear breakup and/or multiple neutron emission. 

Both GDR excitation and BFPP generate a beam of ions with slightly different charge ($Z$) to mass ($A$) ratio, than the circulating ion beam, but with practically unchanged per-nucleon momentum.  These beams gradually diverge from the orbit of the uninteracted ions, and are lost from the beam, reducing the luminosity.   Depending on the beam optics, these beams may remain collimated long enough to strike the accelerator beampipe at a specific location where they might deposit enough energy to quench superconducting magnets or generate radiation damage.  This process is the main factor limiting the LHC luminosity with heavy ion beams \cite{Klein:2000ba,Jowett}.  

In this work, we calculate the cross-section for Coulomb excitation and BFPP for different proposed machine configurations.  We then consider some of the consequences for the proposed designs.


Three different designs are under consideration: two approaches for a U.S.-based, moderate-energy, high-luminosity electron-ion collider (EIC), and the much higher energy CERN LHeC.  The Brookhaven eRHIC and the LHeC designs add electron accelerators to existing hadron colliders, while the Jefferson Laboratory MEIC builds on their existing electron accelerator.   Proposed parameters for the EIC designs are listed in Ref.  \cite{Accardi:2012qut}, while the LHeC is discussed in Ref.  \cite{AbelleiraFernandez:2012cc}.  For $eA$ collisions, the EIC luminosities are presented in terms of electron-nucleon luminosities ({\it i.e.} the number of electron-nucleon collisions), rather than the conventional electron-ion luminosities, which are a factor of $A\approx 200$ lower \cite{Roser}.  

Some key accelerator parameters are listed in Table \ref{tab:accel}.   Recent users group meeting presentations \cite{Litvenko} and a new design study \cite{Aschenauer:2014sua}  have quoted considerably lower luminosities for the eRHIC designs, along with different beam energies, while, recent MEIC presentations \cite{Pilat} quote  $eA$ luminosities up to about 2.5 times higher.  Here, we use the  Ref. \cite{Accardi:2012qut} values as baselines.   At both machines,  higher luminosities are also under discussion in a staged approach, either at a high-luminosity interaction region \cite{Accardi:2012qut} or by machine upgrades.  Eventual electron-nucleon luminosities up to $10^{35}/$cm$^2/$s (electron-nucleus luminosity $5\times10^{32}/$cm$^2/$s) are envisioned \cite{highL,Aschenauer:2014sua}. 

There are several possible LHeC configurations, involving ring-ring and continuous or pulsed ring-linac colliders \cite{AbelleiraFernandez:2012cc}.   Most of the designs have an electron energy of 60 GeV, but a pulsed ring-linac design avoids synchrotron radiation losses, so can reach higher electron energies - 140 GeV.  However, it has a lower luminosity than the other configurations.   Here, we consider a continuous ring-linac configuration, with the `ultimate' per-nucleon $eA$ luminosity from Eq. 6.17 of \cite{AbelleiraFernandez:2012cc}, converted to per-nucleus.  The ring-ring design has the same luminosity.  

 As we will see, the cross-sections do not depend significantly on the collision energy, but the luminosity is critical in determining the overall reaction rate. 

\begin{table}[t]
\begin{tabular}{|l|r|r|r|r|}
\hline
Accelerator & Ion    	& Electron	& Luminosity &  Time  \\
		   & Energy	& Energy		& Peak		& Between \\
		   & (GeV)	 & (GeV)		&(cm$^{-2}$s$^{-1}$) & Collisions \\
\hline
eRHIC   & 100    & 10   & $8.1\times10^{31}$  & 105 ns \\
MEIC    &  40     &   5  & $5.3\times10^{31}$   &  1.25 ns \\
LHeC    & 2940  & 60   & $2.2\times 10^{29}$               & 25 ns \\
\hline 
\end{tabular}
\caption{Parameters for different proposed electron-ion colliders with heavy ion beams.  eRHIC will use gold beams, while the MEIC and LHeC will accelerate lead beams.  The ion energy is per-nucleon, while the luminosity is the $eA$ luminosity, rather than the $e-{\rm nucleon}$ luminosity.   For MEIC, the value for a high luminosity interaction region is used \cite{Accardi:2012qut}.}
\label{tab:accel}
\end{table}

We calculate the cross-sections using the equivalent photon approximation \cite{Bertulani:2005ru}:
\begin{equation}
\sigma = \int \frac{dN}{dk} \sigma_\gamma(k)\  dk
\label{eq:EPA}
\end{equation}
where $k$ is the photon energy in the target nucleus rest frame, $dN/dk$ is the equivalent photon spectrum, and $\sigma_\gamma(k)$ is the cross-section for Coulomb excitation or BFPP.  This approach neglects the dependence of $\sigma_\gamma(k)$ on the photon virtuality ($q^2$).  Usually, this is not a large correction. The integral runs from threshold up to the maximum allowed photon energy.  Since the photon spectrum scales as $1/k$, high photon energies are not important and we will use a maximum energy of $k/10$.  

The photon flux from an electron with energy $E$ is:
\begin{equation}
\frac{dN}{dk} = \frac{\alpha}{\pi k}
 \bigg(1-\frac{k}{E}+\frac {k^2}{2E^2}\bigg)
\ln\big(\frac{q^2_{\rm max}}{q^2_{\rm min}}\big)
\label{eq:fflux}
\end{equation}
where $q^2_{\rm \rm max}$ and $q^2_{\rm min}$ span the $q^2$ range for the photon and $\alpha =1/137.04$ is the fine structure constant.   For an electron with energy $E$ emitting a photon with energy $k$ the kinematic minimum is $q^2_{\rm min} = m_e^2k^2/E(E-k)$ \cite{Budnev:1974de}, 
 where $m_e$ is the electron mass.   

For BFPP, the main contribution is at $q^2\approx k^2$\cite{Bertulani:1997fu}; following Eq. (6.13a) of  \cite{Budnev:1974de}, we use $q_{\rm max}^2=k^2$, so the logarithm in Eq. (\ref{eq:fflux}) is $\ln(E(E-k)/m_e^2)$.

Coulomb excitation can proceed via many different subprocesses.  GDR is the dominant process; it occurs for photon energies from 7-8 MeV (depending on the nucleus) up to about 24 MeV.  At slightly higher energies, additional nuclear excitation channels open up.  These largely lead to multiple neutron emission.  At still higher energies, individual nucleons may be excited, such as to a $\Delta$ resonance.   This work uses the cross-sections from Baltz {\it et al.}  \cite{Baltz:1996as,Baltz:1998ex,Baltz:2002pp,Baltz:2009jk}.   Because of the low photon energies and large cross-sections, GDR is the most important nuclear excitation process.  The GDR resonance is approximated as a Lorentzian, using the parameters in Tab. 3 of Ref. \cite{Veyssiere}.  The cross-section is taken to be zero below the measured 1n threshold of 8.1 MeV (7.4 MeV) for gold (lead).  For the gold data (only) in Ref. \cite{Veyssiere}, a Lorentzian does not appear to be a good fit.  However, newer data has also been fit to a Lorentzian, albeit, for gold, with a slightly smaller (7\%) cross-section, but a wider GDR resonance \cite{Berman:1987zz}.  The two sets of parameters lead to very similar overall $eA$ cross-sections.

For Coulomb excitation, the $q^2$ range depends on the specific sub-reaction.  GDR excitation is a collective nuclear effect with a natural maximum $q_{\rm max}^2 =(\hbar/R_A)^2$, where $R_A\approx 7$ fm is the nuclear radius.  Since this is a looser requirement than $q_{\rm max}^2=k^2$, we use the latter expression. 

Sauter produced a simple analytical approximation to the BFPP cross-section for electron capture to the K-shell.  The Sauter cross-section scales as $Z^5$.  Unfortunately, this approximation is inaccurate for heavy nuclei.   Nuclear correction factors work well at high photon energies \cite{Agger:1997ps}, but cannot reproduce the change in the shape of the cross-section near threshold.  We use numerical data from an exact calculation, the $Z=92$ curve in Fig. 1 of Ref.  \cite{Aste:2007vs}, scaled by $Z^5$ for energies below 14.5 MeV.  At higher energies, we use the Sauter formula with the high$-Z$ correction from Eqs. (9) and (10) of \cite{Aste:2007vs}.  These should agree to within 5\% of the exact cross-sections \cite{Sorensen}.  We increase the BFPP cross-sections by 20\%, to account for electron capture to higher orbitals \cite{Agger:1997ps}.

For the LHeC, there is an additional complication.   Equation \ref{eq:fflux} gives the total photon flux, integrated out to infinite electron-ion impact parameters.  For a given photon energy, the maximum impact parameter is $b_{\rm max}=\Gamma\hbar c/k$, where $\Gamma$ is the Lorentz boost of the electron in the ion rest frame.  For a 1 MeV photon, $b_{\rm max}= 136\ (317) \mu$m for the continuous (pulsed) version, larger than 
the radius of the colliding beams (the heavy ion ring retains the LHC optics, so $R=15.9 \mu$m \cite{PDG}; for LHeC designs with a non-zero beam crossing angle, the effective size may be larger).  The photon flux at larger impact parameters should not be included in the total.  Avoiding the extra flux requires an impact-parameter-sensitive formulation of the photon flux, and some knowledge of the distribution of particles in the beam.    The beam size limitation applies for $k<k_c=8.7$ MeV, so it affects BFPP, but not Coulomb excitation.  For 140 GeV electron beams at the LHeC,  $k_c=20$ MeV, so the beam size also affects Coulomb excitation, but the overall reduction in cross-section is slightly less than 1\%. 

One simple way to estimate the magnitude of the reduction is to set $q_{\rm min} = \hbar/R$ - the transverse momentum uncertainty due to the localization - when $k<k_c$.  This produces a smooth transition at $k=k_c$.  For a 2.5 MeV photon at LHeC-1, the finite beam radius reduces the photon flux by about 6\%.  The overall reduction in cross-section is also about 6\%.   A more detailed calculation is beyond the scope of this paper, but the magnitude of the reduction provides an estimate of the associated uncertainty.    This beam-size effect has been observed at VEPP-4 $e^+e^-$ collider and the HERA $ep$ collider \cite{Kotkin:2003jz}. 

The main uncertainty in the cross-sections is neglect of the photon $q^2$ in $\sigma_\gamma(k)$
One study found an ambiguity of up to a factor of two in the Weizs\"acker-Williams approach due to uncertainties in the $q^2$ cutoffs \cite{Bertulani:1997fu}.  However uncertainties in the $q^2$ range enter only logarithmically in the total cross-section.   At an electron-ion collider, the large Lorentz boosts lead to a very small $q_{\rm min}$; very small $q^2$ will not have a large effect on $\sigma_\gamma(k)$.   Hundley \cite{Hundley:1963eea} found good agreement between the Weizs\"acker-Williams method and a quantum electrodynamics calculation of direct (unbound) pair production.    Overall, the cross-sections should be accurate to 25\%. 

\begin{figure}[htbp]
\begin{center}
\includegraphics[width=0.37\textwidth,angle=270]{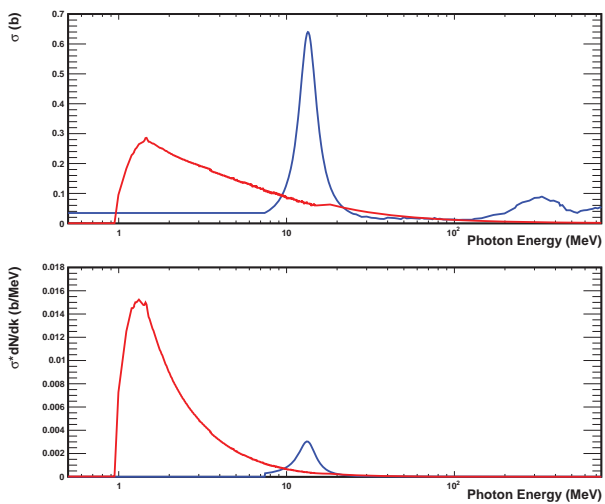}
\caption{(Top) Per-photon cross-section for bound-free (blue) and Coulomb excitation (red) for gold beams at eRHIC, as a function of photon energy, in the target rest frame.  (Bottom)  Flux-weighted cross-sections for $eA$ production of Coulomb excited gold and BFPP.   The photon spectrum scales as $1/k$, so the reactions are  concentrated at low photon energies.   
\label{fig:crosssection}}
\end{center}
\end{figure}

The top panel of  Fig. (\ref{fig:crosssection}) shows the Coulomb excitation and BFPP cross-sections.   The BFPP curve peaks  around 1.4 MeV and decreases slowly at higher energies.  The threshold is less than $2m_e$ because of the electron binding energy.  The slight discontinuity at 14.5 MeV signals the switch from the exact calculation to the Sauter formula.  The Coulomb excitation curve is dominated by the giant dipole resonance.  The second peak, around 300 MeV, is from nucleon excitation to a $\Delta$ resonance.  The bottom panel of the figure shows the cross-sections weighted by the photon flux ($\sigma\times dN/dk$, essentially $\sigma/k$), showing that low-energy photons dominate the cross-sections.  

The 2nd and 3rd rows of Table \ref{tab:sigmax} show the cross-sections for BFPP and Coulex.  The BFPP cross-sections are about 20\% larger than those for Coulex.  Both show only slow variation with collision energy, because the energy dependence enters mainly through the logarithmic term in Eq. (\ref{eq:fflux}).   For BFPP, the LHeC cross-sections are almost identical for 60 and 140 GeV electrons because, at the few-MeV photon energies that dominate the cross-section, the photon flux is constrained by the beam radius, leading to identical logarithmic factors in Eq. (\ref{eq:fflux}).

These cross-sections are quite close to the cross-sections found for the corresponding ultra-peripheral heavy ion interactions, after scaling downward by the square of the ion charge, to account for the reduced photon flux.  For example, at the LHC, the cross-section for BFPP in lead-lead collisions is 281 barns, and that for Coulomb excitation is 220 barns \cite{Baltz:1998ex}.   Scaling downward by $1/82^2$ leads to cross-sections of 41 mb and 34 mb, similar to the 56 and 45 mb for the corresponding $eA$ processes.  The $eA$ cross-sections are of order 50\% larger than the scaled estimates, primarily because of the larger electron Lorentz boost, resulting in a large logarithmic factor in Eq. (\ref{eq:fflux}).  

The 4th and 5th rows of the table show the reaction rates: the Table 1 luminosities multiplied by the cross-sections.  The 6th and 7th rows show the power carried by the beam of altered nuclei.   These beam losses can have consequences for machine and detector operations.   With both processes, the target ion is lost from the beam, decreasing the luminosity.   The neutrons produced from Coulex are also a significant background for experiments, particularly for studies of neutron-free interactions like coherent photoproduction. 

Since the momentum transfer to the ion system is a very small fraction of the total ion momentum, both Coulex and BFPP create a collimated beam of ions with an altered $Z/A$.   These beams deposit energy whereever they hit the beam pipe.  The 6th and 7th rows of the table show the power (in Watts) carried by these beams.  The GDR cross-section is about 2/3 of the total Coulomb excitation cross-section \cite{Baltz:1998ex}.  At RHIC, these ions will lose their collimation ({\it i.e.} spread out)before striking the beampipe, so will distribute their energy around the accelerator ring \cite{Bruce:2007mx}.  At the LHeC, they will strike the beampipe in a well defined location.  Although the power levels are low for the parameters in Tab. 1, they could be significant at a higher luminosity collider. 

Both Coulomb excitation and BFPP remove ions from the beam, and so reduce the luminosity.  In 2014, RHIC ran gold-gold collisions at a center of mass energy of 200 GeV/nucleon, with 111 bunches, each containing $1.6\times10^9$ particles, for a total of $1.78\times10^{11}$ circulating ions \cite{RHIC2014}.  Assuming that eRHIC has similar parameters, a loss rate of $5.5\times10^6$ ions/second (at $8.1\times10^{31}$/cm$^2$/s luminosity) from BFPP plus Coulomb excitation leads to a beam lifetime of 9 hours for one interaction region, or 4 1/2 hours with two interaction points.   The Sept., 2014  eRHIC Design Study, however, anticipated lower beam intensities, $0.6\times10^9$ particles/bunch \cite{Aschenauer:2014sua}.  At these intensities and the Table 1 eRHIC luminosity, the beam lifetimes would be shorter, 200 minutes with 1 IR, or 100 minutes with two IR's.  At the luminosity presented in the design study, the beam lifetime is long enough to avoid trouble. 

Coulomb excitation also produces one or more neutrons, which leave signals in zero degree calorimeters.  The eRHIC neutron production rate of 2.5 Mevents/s should be compared to the beam crossing rate of 9.5 MHz.  Each crossing will contain an average of 1/4 interactions, each with one or more neutrons.  These neutrons are a background contamination for other reactions, particularly for coherent photonuclear interactions where the nucleus remains intact, since the signature for nuclear survival is the absence of neutrons.  The MEIC has a 1.25 nsec interval between collisions, so, the neutron background may be largely removed via the use of a calorimeter with good timing.  
 
\begin{table}[t]
\begin{tabular}{|l|r|r|r|}
\hline
Parameter & eRHIC & MEIC & LHeC \\
\hline
$\sigma$(BFPP) & 37 mb & 39 mb & 56 mb \\
$\sigma$(Coulex) & 31 mb & 28 mb & 45 mb \\
\hline
BFPP Particle/sec      & $3.0\!\times\!10^6$ & $2.1\!\times\!10^6$ & $1.2\!\times\!10^4$ \\ 
Coulex Particles/sec   & $2.5\!\times\!10^6$ & $1.5\!\times\!10^6$ & $1.0\!\times\!10^4$ \\
\hline
BFPP Beam Power & 9.3 W & 2.8 W&  1.1 W\\
GDR Beam Power  & 5.3W& 1.3 W& 0.6 W\\
\hline
\end{tabular}
\caption{Cross-sections and reaction rates for BFPP and Coulex, and the beam power for BFPP
and GDR excitation.}
\label{tab:sigmax}
\end{table}

These issues become problematic at the higher luminosities.  At a luminosity of $5\times10^{32}/$cm$^2$/s with  $0.6\times10^9$ particles/bunch, the eRHIC beam lifetime would be a prohibitively short 34 (17) minutes with 1 (2) IRs.  There would be 1 1/2 neutron-producing-interactions per beam crossing, greatly reducing the efficiency for selecting neutron-free coherent events.  Depending on the machine design, the 50 W of beam power carried by the beam of single-electron ions could be problematic.   It should be noted that fairly high luminosities are required to accomplish the $eA$ physics goals articulated in Ref. \cite{Accardi:2012qut} in a timely manner; most of the $eA$ plans are based on 10 fb$^{-1}/A$ of integrated (per nucleon) luminosity.  

In conclusion, we have considered two processes which will have large cross-sections at proposed electron-ion colliders: bound-free pair production, and Coulomb excitation.   Both processes have small effects on the current accelerator designs, but would present significant obstacles for higher luminosity accelerators.

We thank John Jowett and Tom Roser for help understanding $eA$  luminosities and Peter Jacobs for editorial suggestions.  This material is based upon work supported in part by the U.S. Department of Energy, Office of Science, Office of Nuclear Physics, under contract number DE-AC-76SF00098.

\end{document}